# Exploring the mechanisms of protein folding—The principle of compromise in competition

Ji Xu[a], Mengzhi Han[a,b], Ying Ren[a*], and Jinghai Li[a]

[a]State Key Laboratory of Multiphase Complex System, Institute of Process Engineering, Chinese Academy of Sciences, Beijing 100190, China;

[b]Graduate University of the Chinese Academy of Sciences, Beijing 100039, China;

*To whom correspondence should be addressed. E-mail: yren@home.ipe.ac.cn. Tel.: +86-10-8254 4839.

**Abstract**

Neither of the two prevalent theories, namely thermodynamic stability and kinetic stability, provides a comprehensive understanding of protein folding. The thermodynamic theory is misleading because it assumes that free energy is the exclusive dominant mechanism of protein folding, and attributes the structural transition from one characteristic state to another to energy barriers. Conversely, the concept of kinetic stability overemphasizes dominant mechanisms that are related to kinetic factors. This article explores the stability condition of protein structures from the viewpoint of meso-science, paying attention to the compromise in the competition between minimum free energy and other dominant mechanisms. Based on our study of complex systems, we propose that protein folding is a meso-scale, dissipative, nonlinear and non-equilibrium process that is dominated by the compromise between free energy and other dominant mechanisms such as environmental factors. Consequently, a protein shows dynamic structures, featuring characteristic states that appear alternately and dynamically, only one of which is the state with minimum free energy. To provide evidence for this concept, we analyzed the time series of energetic and structural changes of three simulations of protein folding/unfolding. Our results indicate that thorough consideration of the multiple dynamic characteristic structures generated by multiple mechanisms may be the key to understanding protein folding.

**Introduction**

Proteins are organic macromolecules consisting of twenty different kinds of amino acids and fold into specific spatial conformations, showing dynamic structures that are important for their biological functions. At least three distinct levels of protein structures have been identified: primary structure is the sequence of amino acids, secondary structure refers to regular structures formed by local residues, and tertiary structure is the 3D structure of the protein molecule. The structures of the 20 amino acids are already known in detail, and protein sequences can be identified through experimental techniques like mass spectrometry [1]. However, the resolution of protein structure is comparatively difficult because of the influence of environmental factors such as temperature, pH, and macromolecular concentration. To date, the method used most widely to probe the structure of a protein is to determine its crystal structure through X-ray crystallography. Although most biologists believe that protein structures are static at equilibrium, many scientists have realized that protein structures are dynamic. The dynamic nature and heterogeneity of proteins remain even in the crystalline form because of the high content of solvent in protein crystals [2-4]. In fact, it is logical that dynamic changes are required in proteins for them to perform biological functions. It is acknowledged that proteins exhibit anisotropic motion of individual atoms and collective, large-scale motion over a range of time scales, and that by such dynamic behaviors, proteins can assume a number of almost isoenergetic structures with different conformations [5,6]. Theoretically, a protein molecule can be considered stationary only when there is no thermal motion of any of the atoms, e.g., at 0 K, which would result in a disabled structure without any biological function. Therefore, proteins must be dynamic to perform their biological functions.

The problem of how to obtain a unique structure for a protein, or protein folding, is a classical unsolved problem in life science. As predicted by Li et al. [7,8], the translation of the information contained in the amino acid sequence into the functional structure of a protein is a meso-science problem, and determination of the sequence and stationary 3D structure of proteins are insufficient to reveal their folding/unfolding mechanisms at the meso-scale [9]. Although some techniques have been developed to study the folding process of proteins, like the laser-induced temperature jump [10] and hydrogen exchange [11] methods, it remains difficult to investigate their dynamic behavior. Rapid protein folding events are difficult to capture by current measurement techniques because of their limited spatio-temporal resolution. Meanwhile, computer simulation, especially molecular dynamics (MD) simulation, has become an important tool to investigate the dynamic structures of proteins at atomic resolution [9,12]. Unfortunately, the timescale of current MD simulations is much shorter than the folding time of proteins at room temperature (at least 1 ms). However, the transition state for folding and unfolding is expected to be the same according to the principle of microscopic reversibility [13]. Therefore, MD simulation of protein unfolding is an important tool to study folding mechanisms. Half a century has passed since the problem of understanding protein folding was first recognized. Great theoretical, experimental and computational effort has been expended in attempts to understand the underlying mechanism of protein folding, but a definitive answer still remains elusive.

Historically, two prevalent theories, the thermodynamic control hypothesis and the kinetic control hypothesis, have been proposed. The thermodynamic hypothesis was first proposed by Anfinsen in 1972 [14], and asserts that a native protein in its normal physiological environment is the system with the lowest Gibbs free energy. The experimental evidence used to develop this hypothesis was the observation that the folding/unfolding reactions of many small proteins are reversible [14,15]. During the succeeding 40 years, the thermodynamic hypothesis has been consolidated and led to the concept of a funnel-shaped energy landscape [16,17], although the funnel hypothesis has yet to be validated experimentally. The thermodynamic hypothesis has been widely used in theoretical predictions of the native structures of proteins [18-20] and the study of folding dynamics [21,22]. Although in principle the thermodynamic hypothesis could be tested by comparing all of the possible states found in an exhaustive computer survey of conformational space, such an endeavor is unachievable because of the huge amount of conformational space required, and the rugged topography of the energy surface. An alternative view assumes that the native structure of a protein is kinetically but not thermodynamically stable, and the observed properties should be statistically calculated from a large number of independent MD trajectories [23-25]. This viewpoint holds that there is a high free energy barrier separating the native state from the others, which guarantees the stability of the native structure even if the native state does not correspond to the minimum free energy [23,26-28]. For example, the free energies of the native structures of α-lytic protease [29] and serpins [30] are higher than their respective unfolded structures. The folding of the influenza virus hemagglutinin is induced by low pH [31]. In addition, the β-hairpin folding of the β-switch region of glycoprotein Ibα is induced under flow [32]. The folding of some intrinsically disordered proteins [33] is induced by binding with other molecules like ions, small organic molecules and large biomolecules [34-37]. These findings indicate that the thermodynamic hypothesis is not universal.

According to the accumulated knowledge, in particular, from our understanding of meso-scale phenomena [7], we believe that there must be something missing that is preventing us from understanding the stability condition of proteins. Protein molecules show multi-scale spatio-temporal structures, and the dynamic changes of protein structures are believed to be dissipative and non-equilibrium [9,38]. From the study of complex systems, we believe that a dissipative process is usually dominated by at least two mechanisms [39-41]. This provides a clue that distinguishing between thermodynamic and kinetic stability may not be correct. Protein folding may be dominated by multiple mechanisms, of which free energy is only one. Correspondingly, a protein shows multiple characteristic states dynamically, and minimum free energy is just one of these states. We also think that the concepts of a funnel-shaped energy landscape and its related free energy barrier may be misleading. The switch from one minimum of a landscape to another may be induced by a totally different mechanism to an energy barrier. Surprisingly, this is the same principle behind the dynamic structure of gas-solid flow, which is characterized by alternating existence of a gas-rich dilute state dominated by minimization of energy consumption for transporting and suspending particles and a solid-rich dense state dominated by minimization of gravitational potential with respect to time and space [8,40].

In this paper, we propose that protein folding is dominated by multiple mechanisms, each of which has an extreme tendency that corresponds to a possible characteristic state of the protein, resulting in the dynamic structures of proteins. We provide evidence for this proposal of protein folding by simulations of the folding/unfolding of three protein systems.

**Methods**

**Simulated systems.** Three systems, including folding of RN24, flow-induced unfolding simulations of KID, and thermal unfolding simulations of pKID/KIX complex were simulated in this work. RN24 is constituted of 13 amino-terminal residues of ribonuclease A, and NMR experiments revealed three populated sets of conformations in aqueous solution, including α-helix, partially extended, and bent conformations [42]. KID is a domain from transcription factor CREB that is natively unstructured. NMR spectroscopic analyses revealed that the helical contents in the αA (residues 120-129) and αB regions (residues 134-144) were 50–60% and 10%, respectively, and that helix αB was almost perpendicular to helix αA [37]. Upon binding with KIX, the αB region transforms from coil to helix folding, suggesting that binding of KIX induces considerable conformational changes in KID [33,43]. In this work, the structural transition of KID is studied through unfolding simulations under flow and high temperature.

**Folding simulations of RN24.** Multi-scale simulation is an efficient method to study protein folding. First, the genetic algorithm (GA) method was used to determine the key protein structures corresponding to extreme free energies on the landscape through a global search. Then, hundreds of atomic MD simulations were performed starting from these structures using explicit solvent molecules in the NPT ensemble at 300 K. A comprehensive understanding of folding can be obtained through combination of these two steps. Multi-scale simulations were performed to study the folding of RN24, which is composed of 13 amino-terminal residues of ribonuclease A. Simulation details can be found in our previous work [9].

**Flow-induced unfolding simulations of KID.** The atomic coordinates of KID were obtained from the structure of a pKID-KIX complex (PDB code: 1KDX), which was formed by mutating SEP133 to SER133 by deletion of the phosphate group [43]. The KID was first placed into a water box with dimensions of $10.06 \times 6.04 \times 6.04$ nm$^3$. Three chloride ions were added to neutralize the system. The CHARMM27 force field [44] and TIP3P water model [45] were used for the topology file and parameters, resulting in 35,860 atoms in the system. Particle mesh Ewald summation [46] was used to calculate long-range electrostatic interactions and the van der Waals cut-off radius was 1.4 nm. The system was simulated using an NVT ensemble with $T$=298 K. The stochastic integrator in GROMACS [47] was used in the simulation. First, the energy of the system was minimized using the steepest descent method until convergence to machine precision. This was followed by a 20-ps MD simulation for water relaxation with position constraint applied on KID atoms. Then, to produce constant flow along the x direction, a 0.5-ns simulation was performed. The oxygen atoms of water molecules in the x direction [0, 0.3 nm] of the simulation region were exposed to a constant force of 4.0 pN in the x direction while the KID was fixed. The KID was then allowed to move under flow for 120 ns. The conformations were saved every 50 ps for further analysis.

**Thermal unfolding simulations of pKID/KIX complex.** The protein complex (PDB code: 1KDX) was first placed in a water box with dimensions of $6.5 \times 6.2 \times 6.0$ nm$^3$ containing 7,467 water molecules. N- and C- terminals of both pKID and KIX were chosen as –NH$_2$ and –COOH, respectively. The CHARMM27 force field and TIP3P water model were used for the topology file and parameters, resulting in 24,216 atoms in the system. Particle mesh Ewald summation was used to determine the long-range electrostatic interactions and the van der Waals cut-off radius was 0.9 nm. The energy of the system was then minimized using the steepest descent method until convergence to machine precision. This was followed by a 20-ps isothermal-isobaric MD simulation with T=298 K and P=1.01 atm. Position constraint was applied to the pKID/KIX complex so that the system was relaxed. To maintain the density, the protein unfolding simulation at 498 K was performed in the NVT ensemble. V-rescale temperature coupling in GROMACS was used with two groups, water and protein, coupled separately at a time constant of 0.2 ps. Two 1.0-ms production simulations were performed with different velocity distributions for the atoms at 498 K. The conformations were saved every 10 ps for further analysis.

**Free energy calculation.** The change of Gibbs free energy of a peptide $\Delta G$ can be expressed as the sum of three contributions: the van der Waals interaction $\Delta E_v$, the electrostatic interactions $\Delta E_e$, and the solvation free energy $\Delta G_s$.

$$\Delta G = \Delta E_v + \Delta E_e + \Delta G_s . \qquad (1)$$

For a given structure of a protein, $\Delta E_v$ and $\Delta E_e$ can be calculated using a force field, and the solvation free energy $\Delta G_s$ can be calculated according to the surface properties of the protein [48].

$$\Delta G_s = \sum_i g_i \Delta A_i \quad (2)$$

The kinetically most accessible structures of proteins in the dynamic process can be identified through the landscape of potential of mean force (PMF), which describes how the free energy changes as a function of reaction coordinates. When a trajectory is projected on eigenvectors of its covariance matrix, all structures are fitted to the structure in the eigenvector file, and the projections are called principal components [49]. Here we define the first two principal components of the protein structure, PC1 and PC2, as the reaction coordinates, and the PMF is calculated as [50]:

$$G_X = -RT \ln \frac{N_X}{N_{tot}}, \quad (3)$$

where $T$ is the simulation temperature, $N_X$ indicates the population of state X, and $N_{tot}$ is the total population sampled. Hereafter, $G_X$ is referred to as "energy" for simplicity, which is different from the thermodynamic free energy $G$ in equation (1). A PMF plot is constructed using the same principal components, and the central region of the most populated cluster is chosen as the representative structure.

**Results and Discussion**

The structures and dynamics produced by multi-scale simulations of RN24 folding agreed well with corresponding experimental results [42], and have a much lower computational cost than previous MD simulations [50]. The PMF plot (Fig. 1) was constructed from a set of 51,000 structures sampled using the same principal components. The three most populated conformations were identified, including hairpin, β-sheet and α-helix, which were denoted C1, C2 and C3, respectively. The central region of the most populated cluster was chosen as a representative structure. The C2 structure was found in the most populated region, whereas C3 was found in a shallow, smaller region, and C1 was found in the most shallow region. All of the transitions between α-helix and β-sheet passed through a "coil" conformation indicated by the large azure area surrounding the β-hairpins. Direct transitions between α-helix and β-sheet were never observed in the simulations. The energetic barriers separating C1, C2 and C3 were comparable to the kinetic energy of the atoms of the peptide, implying that transitions between these states occur frequently and can easily be triggered by kinetic perturbation of the environment.

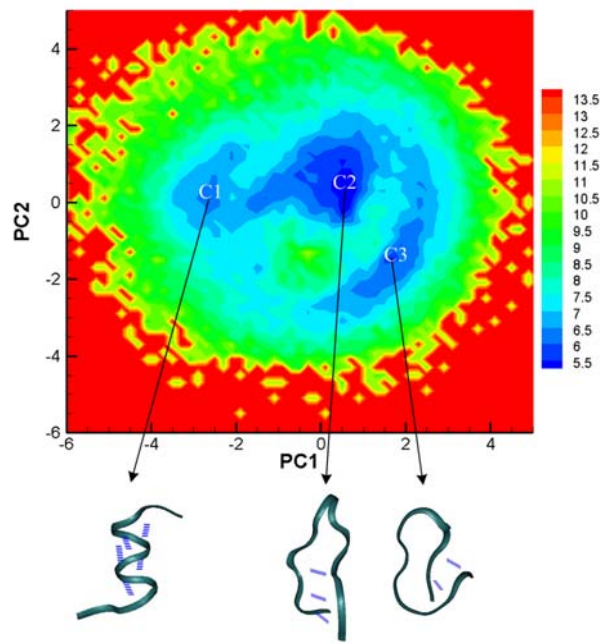

Fig. 1. $G_X$ as a function of the first two principal components [9]. Results are shown in color code formats with a unit of $k_B T$. C1, C2 and C3 indicate three kinetic stable structures. Examples of the most populated conformations are shown as ribbons, with blue dotted lines representing hydrogen bonds.

During the folding process, the free energy profile fluctuates considerably, corresponding to multiple characteristic states that appear alternately and dynamically. Fig. 2 shows the free energy curve of RN24 from a 200-ns trajectory of folding starting from an extended peptide. H1–H3 and L1–L3 denote characteristic structures corresponding to local high and low extreme free energies respectively. The structure corresponding to minimum free energy is just one of these characteristic states. Each structure on the $G$ profiles for all of the 102 folding trajectories can be mapped to a certain point in Fig. 1. Most importantly, further analysis suggests that the characteristic states L2 and L3 belong to the clusters of C1 and C2 in Fig. 1, respectively.

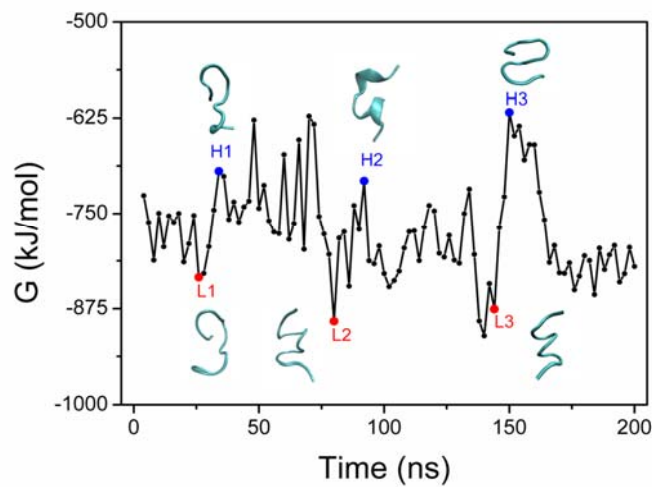

Fig. 2. $G$ as a function of time for RN24 folding from an extended peptide in aqueous solution. Characteristic structures corresponding to extreme energies are shown as ribbons.

Based on hundreds of such free energy profiles starting from different representative peptide structures, a statistical distribution of free energy can be obtained. As shown in Fig. 3, the free energy covers a wide range of ~750 kJ/mol, with a wide but fluctuating peak around -800 kJ/mol. The states of H1–H3 and L1–L3 in Fig. 2 are indicated on the profile. It is not surprising that each state includes several different structures of the peptide, implying that a protein molecule can have several isoenergetic structures. The structures observed in a single dynamic trajectory, as shown in Fig. 2, is only one of the respective ensemble, as shown in the bottom of Fig. 3. The flat top of the profile suggests the existence of multiple states with similar high probability occurring during RN24 folding. This can be further evidenced by the fact that the most populated states C1-C3 identified by the PMF plot in Fig. 1 can be found among the high-probability states like M1, M2 and L1. Another interesting observance is that the characteristic structures H1–H3 and L1–L3 do not exactly correspond to the same extreme in Fig. 3. For example, H1 is a local maximum in Fig. 2, whereas it is located in a local minimum in Fig. 3. Therefore, we think there is some misunderstanding of the free energy theory regarding characteristic states like H1–H3. The traditional theory of a free energy landscape highlights the role of states with minimum energy while assuming there is an "energy barrier" for an energetic transition from one minimum to another. However, we believe these states represent the extreme tendency of another mechanism, and play an equally important role as those with free energy minima during folding. Different structures can be obtained by X-ray analysis of crystallized proteins because multiple characteristic states can form, and each has a probability of being trapped under the same experimental conditions. This is evidenced by the observation of two folding states for the ribosomal protein L20 that coexist in the same crystal unit cell and form under identical physicochemical conditions [51].

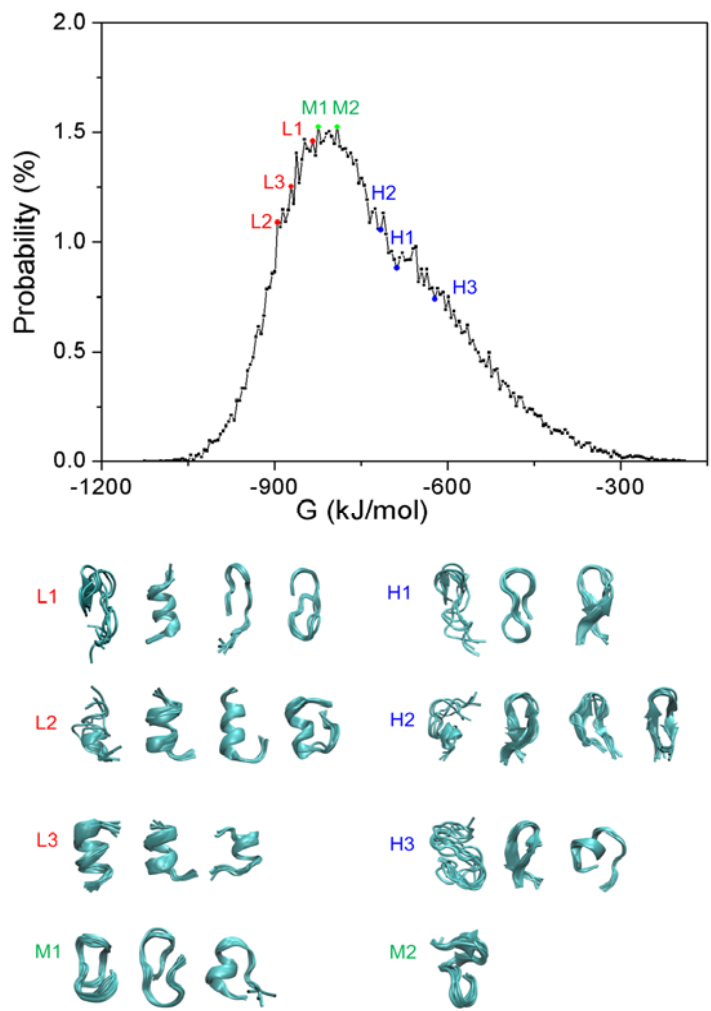

Fig. 3. Probability distribution of free energy for RN24 folding, with the main structure ensemble of the same free energy as H1–H3 and L1–L3 in Fig. 2 shown as ribbons. M1 and M2 indicate two states with the largest possibilities.

For flow-induced unfolding of KID, as shown in the PMF plot in Fig. 4, starting from the structure of KID with both αA and αB in helical conformation, the structure gradually evolved to the four highly populated conformations, C1, C2, C3 and C4. The protein structures of the centroids of these four states are shown as ribbons in Fig. 4, and considerable differences were observed in the helical content and relative position between the regions of αA and αB. The free energy profile as a function of time determined for this system (Fig. 5(a)) suggests that αB gradually transits from helix to random coil while most of αA remains in helical conformation, which agrees with the structure of KID in free solution [37] and can be considered as the inverse dynamics of the structural changes of an unstructured KID upon binding to KIX. The free energy profile fluctuates substantially over time, exhibiting multiple characteristic states of the protein such as H1–H3 and L1–L3. These dynamic structures corresponding to high and low extreme free energies, respectively, appear alternately, implying that there should be at least one other mechanism that compromises and competes with the free energy mechanism. A diagram relating the root mean square distance (RMSD) between the dynamic structures shown in Fig. 5 to the PMF plot in Fig. 4 indicates that H1 and H2 belong to the kinetically most accessible states C2 and C3 respectively. This suggests that structures with high free energies can be stabilized under flow. That is, protein folding is a dynamic process between free energy and at least one other dominant mechanism. Fig. 5(b) further suggests that the structures corresponding to free energy minima or maxima are only representative states, representing the extreme tendency of free energy or other dominant mechanisms. Those with higher free energy, like H1-H3, do not possibly relate with the free energy barriers, as explained by the free energy funnel theory.

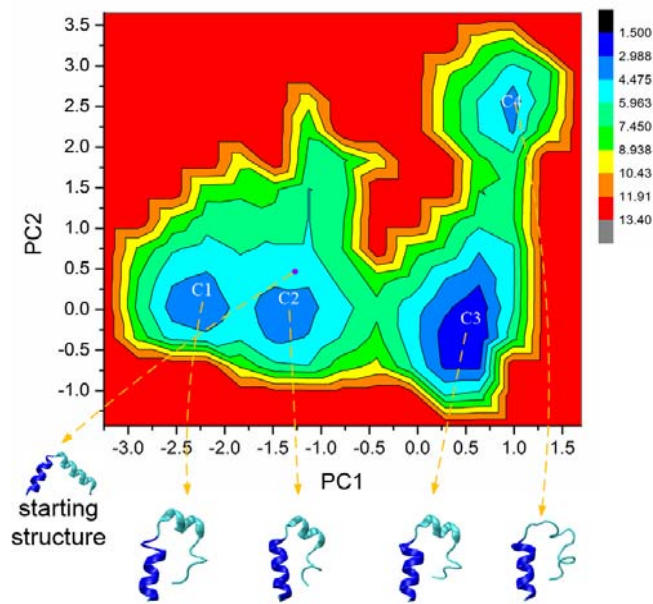

Fig. 4. $G_X$ as a function of the first two principal components. Results are shown in color code format with the unit of $k_B T$. C1, C2, C3 and C4 are four kinetically stable structures. Examples of the most populated conformations are shown as ribbons. αA is shown in blue and other residues are shown in cyan.

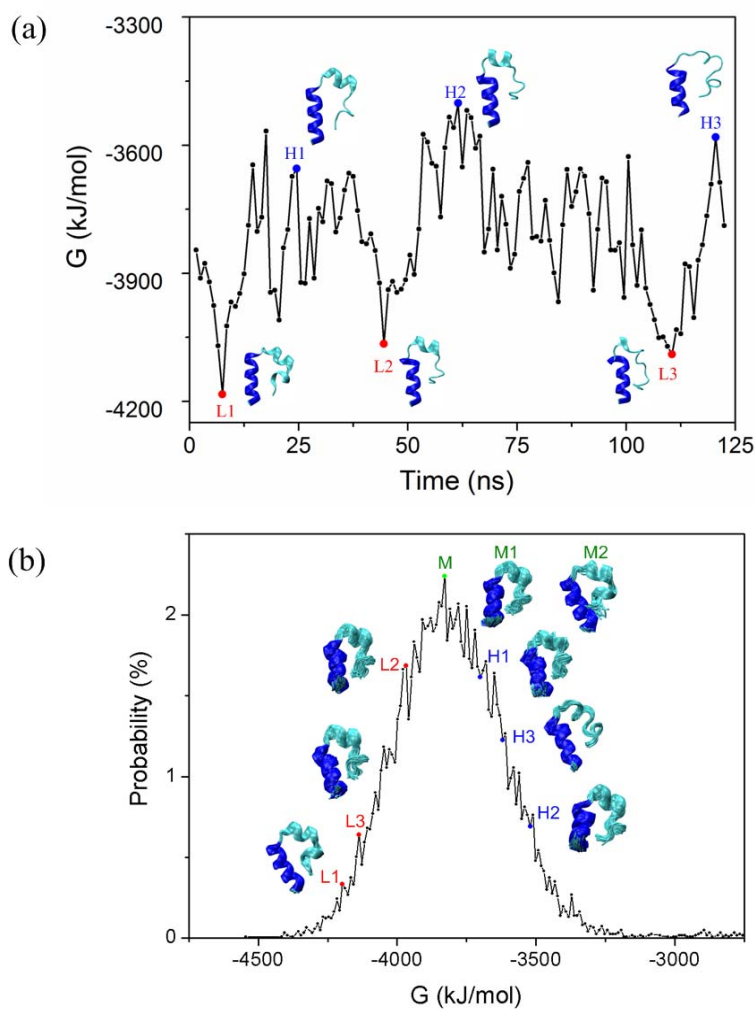

Fig. 5. (a) *G* as a function of time for KID under flow, with characteristic structures corresponding to extreme energies shown as ribbons. (b) Probability distribution of free energy, with the main structure ensemble of the same free energy as L1–L3 and H1–H3 shown as ribbons. M can be divided into two clusters, M1 and M2.

The PMF plot of the structural changes of pKID during thermal unfolding of the pKID/KIX complex is presented in Fig. 6. This plot was constructed with all of the 200,000 conformations generated from the two 1.0-ms simulations. It shows three most populated clusters, with C4 quite similar to the structured pKID with most αA and αB regions in helices. Further analysis of the structures in these clusters suggests that two distinct structures, C1 and C2, are located in the same region of the PMF plot. The backbone RMSD between the average structure of C1 and C2 was 0.275 nm, which mainly results from structural differences in the random coil parts between αA and αB with the residues from ILE127 to ARG132. This may be attributed to the large number of structures occupying a large continuous region at the bottom left of the PMF plot. These structures are similar in that both αA and αB are partially unfolded and linked by a random coil, resulting in different orientations of the backbone of pKID. This further implies that the common method to identify structures using the PMF plot is not precise enough to distinguish such conformations when a cluster of the most populated structures occupies a comparably large space. In this case, another classification method with higher resolution should be used together with the PMF plot. The protein structures of the centroids of C1–C4 are shown as ribbons in Fig. 6. There are considerable differences in the helical content in the αB region and the relative positions of αA and αB. The large continuous region in dark blue suggests easy transitions among these structures.

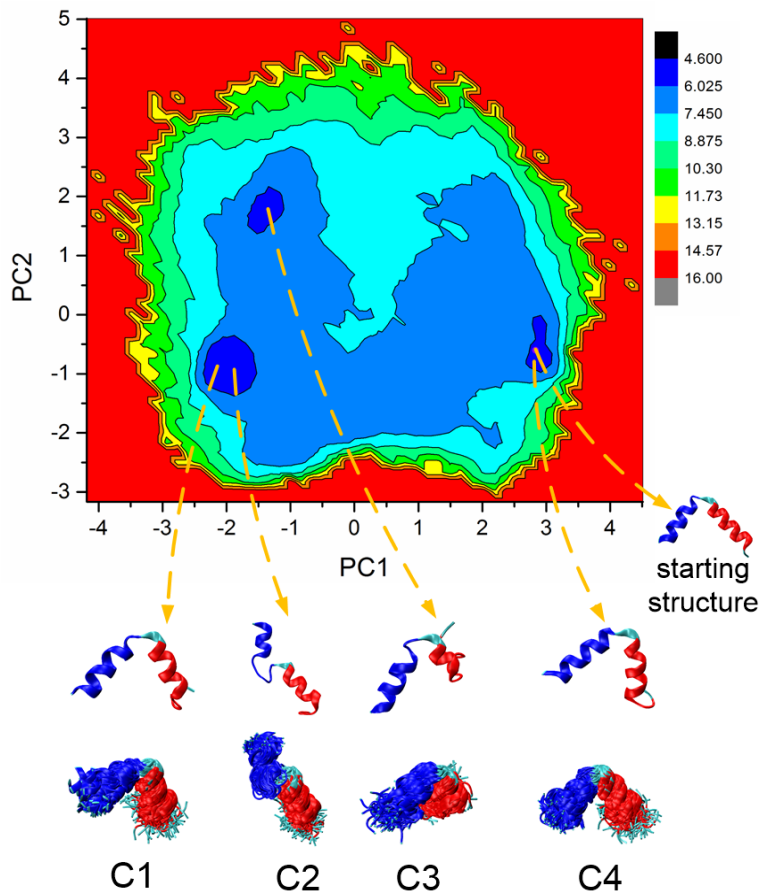

Fig. 6. $G_X$ as a function of the first two principal components. Results are shown in color code formats with the unit of $k_B T$. C1, C2, C3 and C4 are four kinetically stable structures. Examples of the most populated conformations are shown as ribbons, with the two rows indicating the average structure and all of the structures in clusters C1–C4, respectively. αA and αB are shown in blue and red, respectively, while other residues are shown in cyan. The bottom images are drawn from all of the structures in each respective cluster.

For such a complex system, the free energy can be classified into the intraprotein free energy of pKID and the binding free energy of pKID to KIX, indicating the structural changes caused by the interactions within pKID itself and from binding to KIX, respectively. For each 1.0-ms simulation, 100,000 conformations were obtained. To clarify the evolution of free energy, we used a cluster method. The clusters were defined according to the RMSD for all of the atoms in pKID. Starting from the first conformation, if the RMSD of the following structure is within 0.1 nm of the mean RMSD of all of the previous structures in the cluster, then it is assigned to the same cluster, otherwise a cluster with a new conformation begins.

The energy evolution of clusters in the first 180 ns of the first simulation is plotted in Fig. 7(a). The upper panel of Fig. 7(a) suggests that there is a compromise between the two interactions.

That is, the two mechanisms alternatively dominate the dynamic process, resulting in considerable fluctuations in each curve. For the complex of pKID/KIX, the free energy profile shows less fluctuation, but still contains many free energy extremes. Some representative structures of pKID corresponding to extreme energies are shown as ribbons in the bottom panel of Fig. 7(a) to offer an intuitive view of the unfolding dynamics. The free energy profile as a function of time suggests that αA transits among different conformations, such as partial α-helix (H1), almost completely α-helix (L1), and random coil (L2). Meanwhile, αB always remains in helical conformation, which agrees with other simulation results [52] and experimental results for the pKID/KIX complex showing that the interactions between αB and KIX are stronger than that between αA and KIX [53]. However, the PMF plots of KID unfolding (Fig. 4) and pKID/KIX unfolding (Fig. 6) show a strong helical tendency for the αA region. Therefore, the structures with αA almost completely α-helix, such as L1, are important in the evolution of the structure of pKID. In addition, the conformations containing other structures of αA, are also encountered in the unfolding dynamics. This suggests that the conformations that do not belong to the kinetically most accessible structures are equivalently important in structure evolution. If the hypothesis that folding is the reverse of unfolding [54,55] holds for this system, folding of pKID should be a dynamic process under the control of free energy itself and the binding energy from protein KIX. The distribution profile of free energy in Fig. 7(b) is similar to that in Fig. 5(b) except that it is more compressed. The state with the most probability, M, includes two different clusters, M1 and M2. Each cluster agrees with a state in the most populated cluster in the PMF plot (Fig. 6): C1 and C2, respectively.

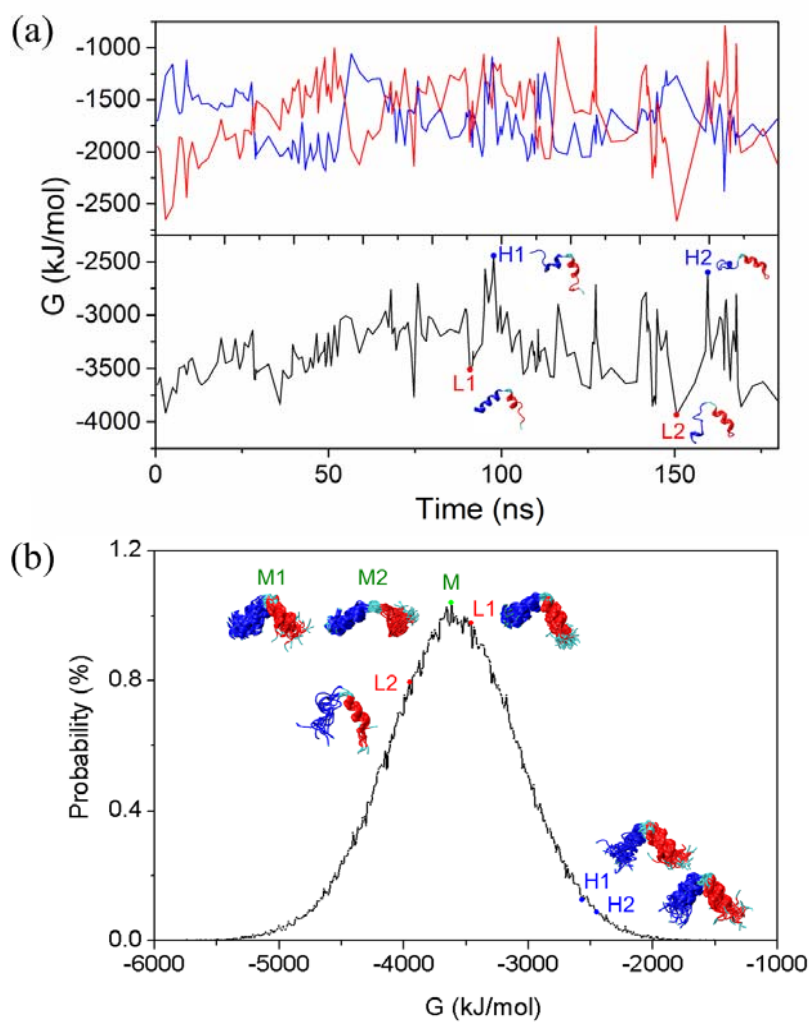

Fig. 7. (a) Upper panel: intraprotein free energy of pKID (blue) and binding free energy of KIX to pKID (red) as a function of time. Lower panel: sum of intraprotein and binding free energy as a function of time, with characteristic structures corresponding to extreme energies shown as ribbons. (b) Probability distribution of free energy, with the main structure ensemble of the same free energy as L1, L2, H1 and H2 shown as ribbons with the same color scheme as Fig. 6. M can be divided into two clusters, M1 and M2.

Based on the above discussion, the structures of proteins show multiple populated conformations corresponding to representative dominant mechanisms. A concept that governs the meso-scale structures in complex systems is the underlying principle of compromise, which can be described as compromise in the competition between different dominant mechanisms. That is, mechanisms compromise with each other to reach a relative overall extremum, resulting in spatio-temporal multi-scale structures in complex systems [56]. As shown in Fig. 8(a), in gas-particle systems, multiple characteristic states exist that correspond to extreme drag force, which can be attributed to the spatial and temporal compromise of the movement tendencies of the fluid and solid phases [57]. This principle is also applicable to protein structures. As shown in Fig. 8(b), the free energy of a protein can be treated as dominant mechanism A, and additional mechanisms dominating protein folding can be environmental factors like temperature, pressure, pH, and interactions with surrounding biomolecules. Taking the folding of RN24 as an example, the additional mechanism can be summarized as the kinetic perturbation from the environment. The compromise between two dominant mechanisms results in alternative realization of protein structures with low free energy such as α-helices and β-sheets, as well as random coils with high free energy.

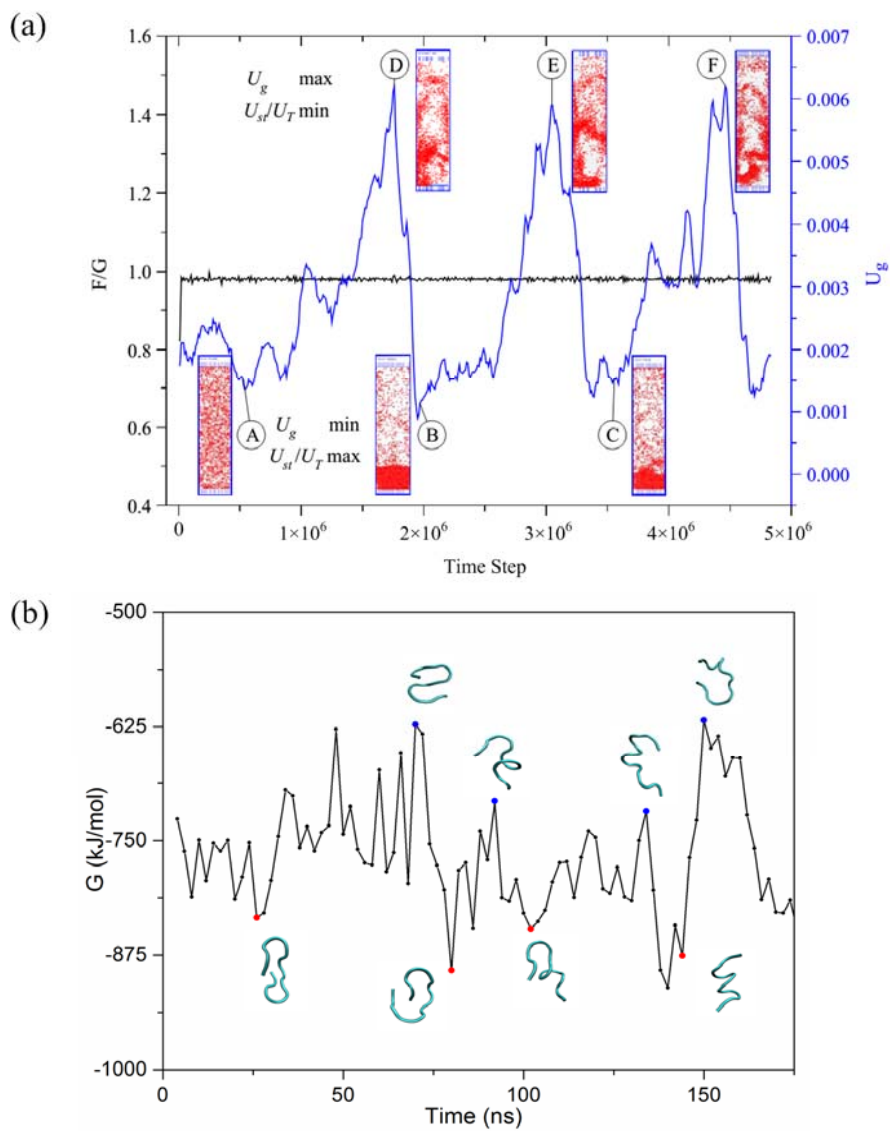

Fig. 8. Similarity between the complex systems of fluidization and protein folding. (a) Temporal variation of the ratio of drag force to gravity ($F/G$) and gas flow velocity ($U_g$) in a particle-fluid system, including some characteristic heterogeneous structures [57]. (b) Temporal variation of the free energy for RN24 folding in aqueous solution, with some characteristic structures shown as ribbons.

Protein folding *in vivo* is in an extremely complex environment, with high macromolecular content, and differences in flow, pressure, temperature, and pH between different organelles. Therefore, additional dominant factors should be explored to elucidate the mechanisms of protein folding and stability in biological systems. As shown in Fig. 9, the alternative realization of the extreme tendencies of mechanisms A and B results in multiple dynamic conformations during protein folding. However, this realization does not occur simultaneously, as evidenced by the time series of free energies of the system. Although the importance of free energy minima has been increasingly recognized and even overemphasized [17], other characteristic states of proteins, like those dominated by mechanism B, are always treated as transient structures explained by the energetic barriers in the theory of free energy landscapes, and are thus ignored. We believe that the structures dominated by mechanism B also exist widely as characteristic states in protein systems, and should be given more attention because they have equal importance during protein folding and functioning as those dominated by mechanism A. Thorough consideration of the multiple dynamic characteristic structures of proteins being dominated by multiple mechanisms shows promise to reveal the underlying mechanisms of protein folding. This may be the key to resolving the problem of understanding protein folding.

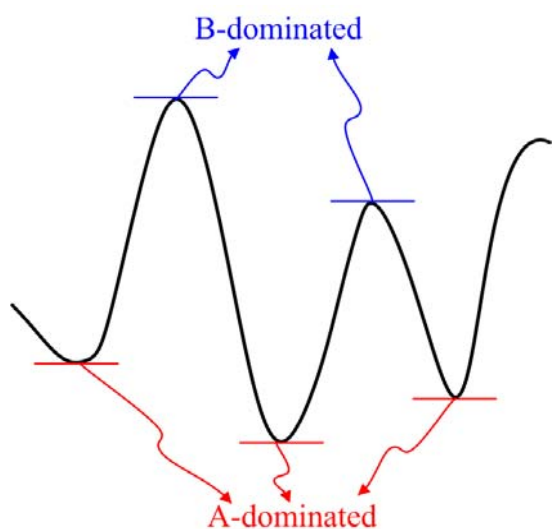

Fig. 9. Schematic diagram showing the free energy landscape and the extreme tendencies of different dominant mechanisms.

**Conclusions**

We have shown that protein folding is a meso-science issue. Determination of the 3D structure of a protein from its amino acid sequence requires a thorough understanding of the dynamic structures generated by the nonlinear non-equilibrium folding process. Based on our understanding of complex systems, the minimum free energy criterion is possibly not the exclusive criterion governing the stability of proteins. We believe that other dominant mechanisms compromise with free energy to shape the dynamic structures of proteins. The concept of additional dominant mechanisms should be explored further through the study of more protein systems with different physicochemical properties and in different environments. Our main ideas of protein folding can be summarized as follows:

(1) Structures of proteins are dynamic, showing multiple characteristic states alternately, of which the native state characterized by free energy minimum is just one.
(2) Protein folding is controlled by multiple dominant mechanisms including the free energy mechanism, and each corresponds to a possible characteristic state of the protein.
(3) Distinguishing between the thermodynamic and kinetic stability of a protein may not be reasonable. Thermodynamic stability overemphasizes the dominating mechanism of free energy, while kinetic stability overlooks the effect of free energy but overemphasizes that of the mechanism competing against free energy.

(4) The dynamic process of protein folding should be depicted through the time series of both its energetic and structural changes, instead of the topology of free energy landscape. The switch from one characteristic state to another should result from the dominance of another mechanism rather than an energetic barrier in the landscape.

Because it is difficult to measure the dynamic structures of proteins experimentally, computations are very important in exploring the underlying mechanisms of protein folding. A promising solution is to develop a computational paradigm that demonstrates structural consistency between the folding problem, modeling, software, and hardware to raise the efficiency and capability of supercomputing [56]. With the combined effort of computation, theory and experiment, the underlying mechanisms of protein folding should finally be unveiled.

**Acknowledgements**

This work was supported by the National Natural Science Foundation of China under Grant No. 21103195, and the Knowledge Innovation Program of Chinese Academy of Sciences under Grant No. KGCX2-YW-124.